\documentclass{optica-article}
\journal{opticajournal} 

\articletype{Research Article}
\usepackage{lineno}

\begin{document}

\title{Dual-comb mode-locked Yb:CALGO laser based on cavity-shared configuration with separated end mirrors}

\author{Ruixin Tang \authormark{1,2}, Ziyu Luo \authormark{1,2}, Pengfei Li \authormark{1,2}, Pengrun Ying \authormark{1,2}, Haiyang Xie \authormark{1,2}, Siyuan Xu \authormark{1,2}, Hui Liu \authormark{1,2,3,*}, Jintao Bai \authormark{1,2,4,*}}

\address{\authormark{1}State Key Laboratory of Energy Photon-Technology in Western China, Northwest University, Xi’an 710127, China\\
\authormark{2}Institute of Photonics and Photon-Technology, Northwest University, Xi’an 710127, China\\}

\email{\authormark{3}liuhui$\_$gzs@nwu.edu.cn} 
\email{\authormark{4}baijt@nwu.edu.cn} 


\begin{abstract*} 
Dual-comb spectroscopy typically requires the utilization of two independent and phase-locked femtosecond lasers, resulting in a complex and expensive system that hinders its industrial applications. Single-cavity dual-comb lasers are considered as one of the primary solution to simplify the system. However, controlling the crucial parameter of difference in repetition rates remains challenging.  In this study, we present a dual-comb mode-locked Yb:CALGO laser based on a cavity-shared configuration with separated end mirrors. We employ two pairs of end mirrors and two thin-film polarizers angled at 45 degrees to the cavity axis, leading to separating the cross-polarized laser modes. We achieve simultaneous operation of two combs at approximately 1040 nm with pulse durations of around 400 fs and an average power exceeding 1 W. The repetition rates are approximately 59 MHz and their difference can be easily tuned from zero up to the MHz range. By effectively canceling out common mode noises, we observe minimal fluctuation in the repetition rate difference with a standard deviation of about 1.9 Hz over ten minutes, while experiencing fluctuations in repetition rates as large as 90 Hz. We demonstrate the capabilities of this system by utilizing the free-running dual-comb setup for asynchronous optical sampling on a saturable absorber and measuring etalon transmission spectrum. This system allows for simple and independent control of the repetition rates and their difference during operation, facilitating the selection of optimal repetition rate difference and implementation of phase-locking loops. This advancement paves the way for the development of simple yet high-performance dual-comb laser sources. 

\end{abstract*}

\section{Introduction}
Dual-comb laser (DCL), comprising two coherent optical frequency combs with slightly different repetition rates, has found extensive applications. Among these applications, the most notable one is the dual-comb spectroscopy (DCS)\cite{1}, encompassing ultrabroadband IR spectroscopy~\cite{2,3,4,5}, mid-infrared near-field microscopy~\cite{6}, spectrally resolved Lidar~\cite{7},  coherent Raman spectroscopy~\cite{8,9,10},  THz spectroscopy~\cite{11,12}, and optomechanical spectroscopy\cite{13}. When employed in pump-probe experiments, the DCL will fully unleash its potential for precise material characterization and measurement, contributing to advancements in THz time-domain spectroscopy\cite{14,15}, ultrafast photoacoustics~\cite{16,17}, and ultrafast magnetization dynamics~\cite{18,19} . The field of laser ranging has greatly benefited from the utilization of DCL~\cite{20,21,22}, which simultaneously offers high precision and extended ambiguity range. 

Conventional dual-comb lasers typically operate using two independent and phase-locked optical frequency combs (OFCs)~\cite{23,24}. However, the high cost, complex design, and large scale of such systems impede their widespread adoption for practical applications. Therefore, there is an urgent need for a cost-effective, simplified, and compact dual-comb laser (DCL). One promising approach to achieve this is through a single-cavity or quasi-single-cavity DCL that sustains two OFCs with different repetition frequencies within a stringent or non-stringent shared resonator~\cite{25,26,27,28}. By canceling out common mode noises between the OFCs in the shared cavity, desirable mutual coherence can be achieved without requiring a phase-locking device. As a result, single-cavity DCLs have garnered significant attention. Although the concept of single-cavity DCLs emerged as early as in the 1990s to synchronize pulses with different wavelengths in modelocked lasers~\cite{29,30}, it is only in recent years that these devices have experienced rapid development with emphasis on asynchronous operation and high mutual coherence characteristics.

In 2015, the U. Keller research team achieved the realization of a single-cavity dual-comb laser in a modelocked integrated external-cavity surface emitting laser (MIXSEL) by incorporating a birefringent crystal into the cavity, thereby introducing the concept of dual-comb modelocked laser~\cite{31}. In 2017, the same team demonstrated the feasibility of DCS using a free-running dual-comb MIXSEL, significantly reducing the complexity of DCS laser systems~\cite{32}. The application of this concept was later extended to a mode-locked Yb:CaF$_2$ laser~\cite{33}. In 2022, they realized a dual-comb mode-locked laser using spatially multiplexed all-solid-state lasers~\cite{34}. T. Ideguchi and his colleague reported in 2016 on a Ti:sapphire bidirectional ring dual-comb mode-locked laser utilizing a ring cavity structure and showcased its potential for broadband DCS applications~\cite{35}. This bidirectional ring approach was subsequently adopted by other groups for fiber ring lasers as well~\cite{36,37,38,39,40}. Currently, various approaches have been explored to achieve single-cavity dual-comb lasers by exploiting the degrees of freedom in polarization, spatial mode, wavelength, and circular direction. These strategies have demonstrated successful applications both in solid-state and fiber lasers\cite{RN835}.

Despite these advancements, achieving flexible control over the repetition rate difference (RRD) remains a challenge for single-cavity DCLs. In the case of all-solid-state single-cavity dual-comb mode-locked lasers, adjusting the RRD typically necessitates realigning the cavities or modification of other parameters rather than simply translating a mirror~\cite{26,33}. Consequently, this approach offers only a very limited range of RRD control. Moreover, due to the lack of independent control over repetition rates, enhancing coherence or mutual coherence in single-cavity DCLs beyond that achieved under free-running conditions poses a significant challenge. Therefore, single-cavity DCLs in free-running generally exhibit moderate levels of mutual coherence. A current trend in the field is the establishment of a dual-comb laser system featuring a shared cavity and enhanced control over RRD. This has been recently demonstrated using thin-disk lasers\cite{27,28} and fiber lasers\cite{RN877}. 

In this study, we present an all-solid-state dual-comb mode-locked laser based on a cavity-shared configuration with separated end mirrors, enabling flexible control of the RRD. This dual-comb laser utilizes a bulk Yb:CALGO crystal and generates each comb with an average power exceeding 1 W and a pulse duration of approximately 400 fs. The two combs share all intracavity elements but employ separate end mirrors. This design facilitates extensive tuning of the RRD during mode-locking operation by longitudinally adjusting the position of the end mirrors using a three-dimensional (3D) translation stage, allowing for optimal and flexible selection of this crucial parameter. With fundamental repetition rates approaching 59 MHz, the RRD can be easily tuned from zero up to the MHz range. The RRD exhibits minimal fluctuations, with a standard deviation of only 1.9 Hz over a period of 10 minutes, while the fluctuation magnitude in repetition rate for each comb is approximately 90 Hz. We demonstrate the spectroscopic measurement capability of this dual-comb system in free-running mode by measuring the transmission spectrum of an etalon. Furthermore, under free-running conditions, we showcase its usability in pump-probe experiments through asynchronous optical sampling (ASOPS) on a semiconductor saturable absorber mirror  (SESAM) which enables rapid resolution of absorption dynamics on picosecond timescales. If higher precision measurements are required, phase-locking can be independently applied to each comb within this co-cavity system to enhance coherence or mutual coherence.

\section{Experimental setup}
The dual-comb mode-locked laser, as depicted in Fig. 1, is constructed using a bulk Yb:CALGO crystal (CASTEC) and optically pumped by a 50 W fiber-coupled diode laser operating at 981 nm. The crystal has been doped with 3 atomic percent of ytterbium ions and is c-cut with dimensions of 3 mm × 3 mm × 3 mm. The pump beam passes through a telescope consisting of two lenses (75 mm × 150 mm) before being focused into the gain medium. To achieve dual-comb operation, a birefringent crystal with a thickness of 2 mm (cut at an angle of 45° with respect to the optical axis) is inserted after the fiber end (see setup in Fig. 1 inset), resulting in the splitting of the pump single beam into two beams that walk off transversely from each other. As a result, there are two spatially separated pump spots within the gain medium, with a separation distance of 200 $\upmu$m. It should be noted that this spatial separation of pumping plays a crucial role in enabling dual-comb operation; otherwise, only one comb output would occur due to gain competition.

\begin{figure}[ht!]
	\centering\includegraphics[width=12cm]{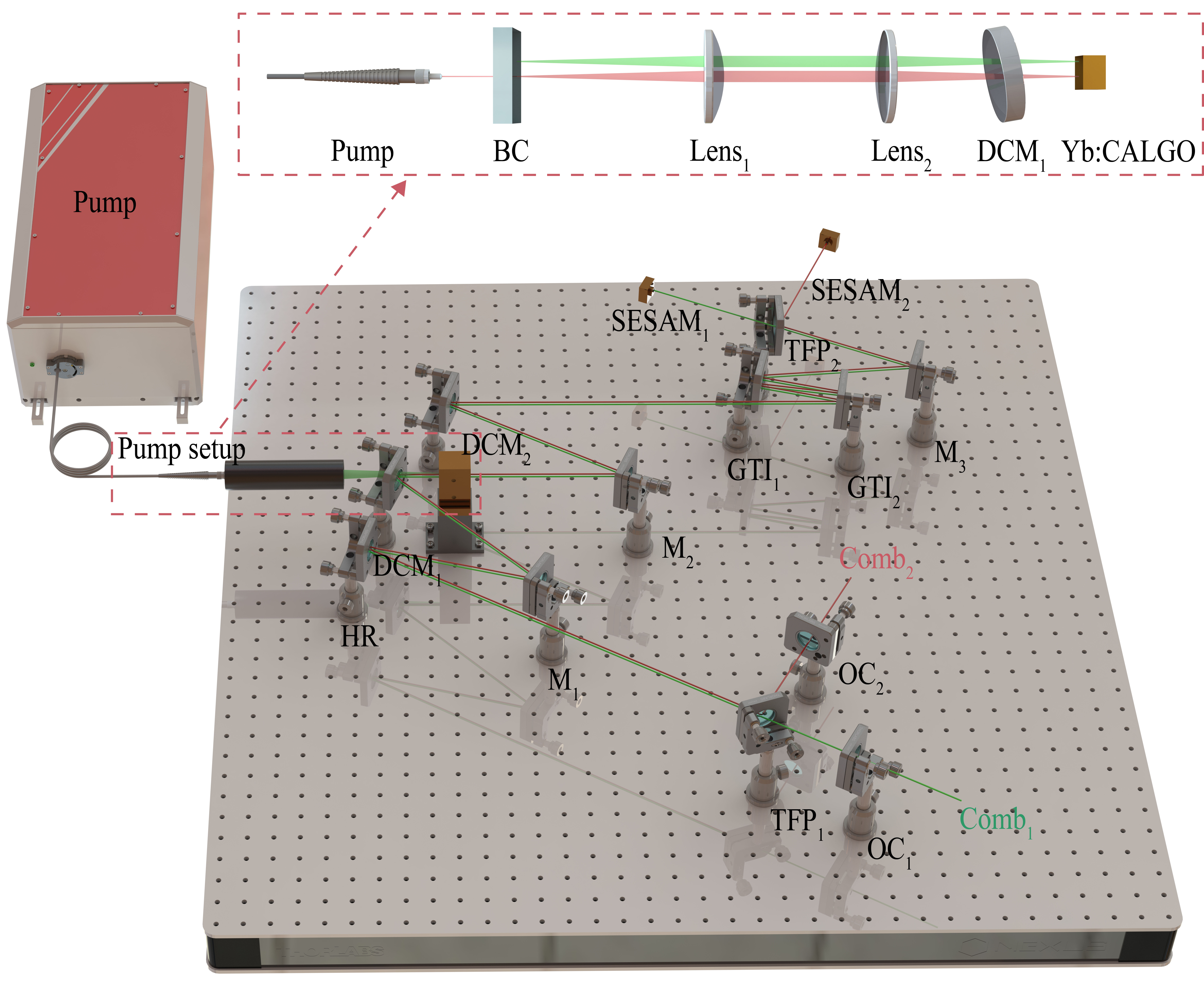}
	\caption{Schematic drawing of the dual-comb laser system. SESAM, semiconductor saturable absorber mirror; TFP, thin-film polarizer; DCM, dichroic mirror; HR, high reflection; M, mirror; OC, output coupler; GTI, Gires–Tournois interferometer .}
\end{figure}

To achieve soliton mode-locking in the negative dispersion regime, dispersive mirrors (GTI$_1$, GTI$_2$, M$_1$ and M$_2$) are introduced, yielding a net group delay dispersion of $-$5100 fs$^2$ per round trip. The concave mirrors M$_1$–M$_3$ have curvature radii of 500 mm, 300 mm, and 500 mm respectively. The DCMs represent dichroic mirrors with high transmission in the wavelength range of 900–1000 nm and high reflection (HR) in the range of 1030-1050 nm. The flat output couplers (OCs) exhibit a transmittance of 2.4$\%$. For self-starting and maintaining mode-locking, we employed a commercial SESAM (BATOP) with a modulation depth of 1.2$\%$ and saturation fluence Fsat = 60 $\upmu$J/cm$^2$. To adjust the distance between M$_3$  and the SESAMs, we mount the SESAMs on micrometer-driven translation stages.

The dual-comb operation is achieved by the arrangement of two nearly identical laser resonators, which share almost all cavity mirrors except for the SESAM and OC. In the cavity, two 3-mm thin-film polarizers (TFP) (Edmund optics) are positioned at 45° angle to enable polarization selection of the laser modes. The TFP reflects s-polarization and transmits p-polarization. It should be noted that simultaneous oscillation of cross-polarized modes with balanced power is only possible when the crystal is c-cut, not if it is a-cut. By separating the SESAMs and OCs, independent tuning of repetition rates for each comb becomes possible. With this configuration, RRD can be easily tuned from zero up to the MHz range by adjusting the end mirrors' position. This provides a significant advantage over solid-state single-cavity DCLs that rely on spatially multiplexed or birefringent or bidirectional cavities, where tunability of RRD is quite limited and typically requires realignment of cavities or modification of other parameters rather than simple mirror translation.

\section{Dual-comb characteristics}
\begin{figure}[ht!]
	\centering\includegraphics[width=13cm]{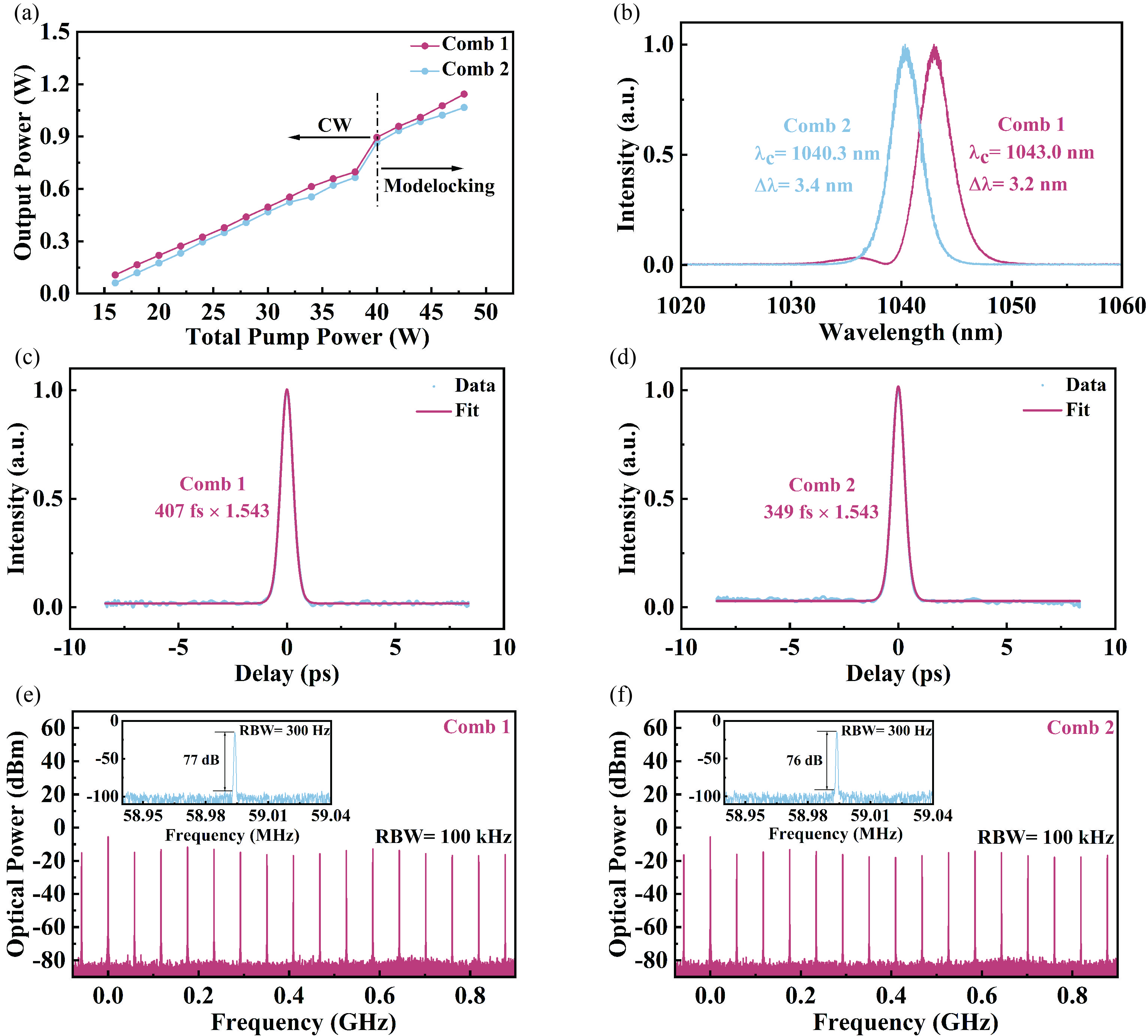}
	\caption{ The performance evaluation of the dual-comb. (a) Laser output power versus total pump power, (b) Optical spectra of comb 1 (blue) and comb 2 (red), (c)-(d) Autocorrelation trace with sech$^2$ fit, (e)-(f) Laser output dected by a fast photodiode and measured by an electrical spectrum analyzer where the insets indicate signal-to-noise ratio with a resolution bandwidth (RBW) of 300 Hz.}
\end{figure}

The output laser properties were characterized with self-started and stable mode-locking of both combs. Here, we designate the two combs as comb 1 (p-polarized output) and comb 2 (s-polarized output). Figure 2(a) displays the individual comb output power versus the total incident pump power, with maximum average output powers of 1.07 W and 1.14 W for the two combs, respectively. The optical spectra were measured using an optical spectrum analyzer (OSA, Yokogawa, AQ6370B), as shown in Fig. 2(b), revealing a full-width at half-maximum (FWHM) bandwidth of 3.2 nm (3.4 nm) at wavelength of 1043 nm (1040 nm). The pulse autocorrelation traces were obtained by an intensity autocorrelator (APE, pulseCheck) at the highest output power level, yielding a pulse duration of 407 fs (349 fs), as depicted in Fig. 2(c)-2(d).  After capturing the pulse sequences with a high-speed InGaAs detector (Thorlabs, DET08CL/M), the repetition rates and harmonics were measured using an electrical signal analyzer (ESA, Keysight, N9000A). A signal to noise ratio of approximately 77 dB (76 dB), as shown in Fig. 2(e)-2(f), indicates stable fundamental modelocking at a repetition rate of ~59 MHz for both combs. The difference between the combs in these characteristics is mainly attributed to the weak polarization anisotropy within the gain crystal.

To demonstrate the high mutual coherence between the two combs, we simultaneously measured their repetition rates. The repetition rates were accurately determined using two counters (Stanford Research System, SR620) with a precision of millihertz. Figure 3(a) shows the temporal evolution of the two repetition frequencies and their difference. The repetition rates of both lasers exhibit drifts of approximately 90 Hz; however, the difference between them shows only weak fluctuations with a standard deviation of 1.9 Hz over a duration of 10 minutes. This observation confirms the remarkable suppression of common noises within this dual-comb laser system. The longitudinal positions of the SESAMs can be continuously adjusted with 3D translation stages. In this study, the translation of the SESAMs allows tuning the RRD between the two combs from 0 to 1.2 MHz, as shown in Fig. 3(b), with a negligible loss on the laser output performance. Additionally, control over RRD can also be achieved by manipulating the OCs.
\begin{figure}[ht!]
	\centering\includegraphics[width=13cm]{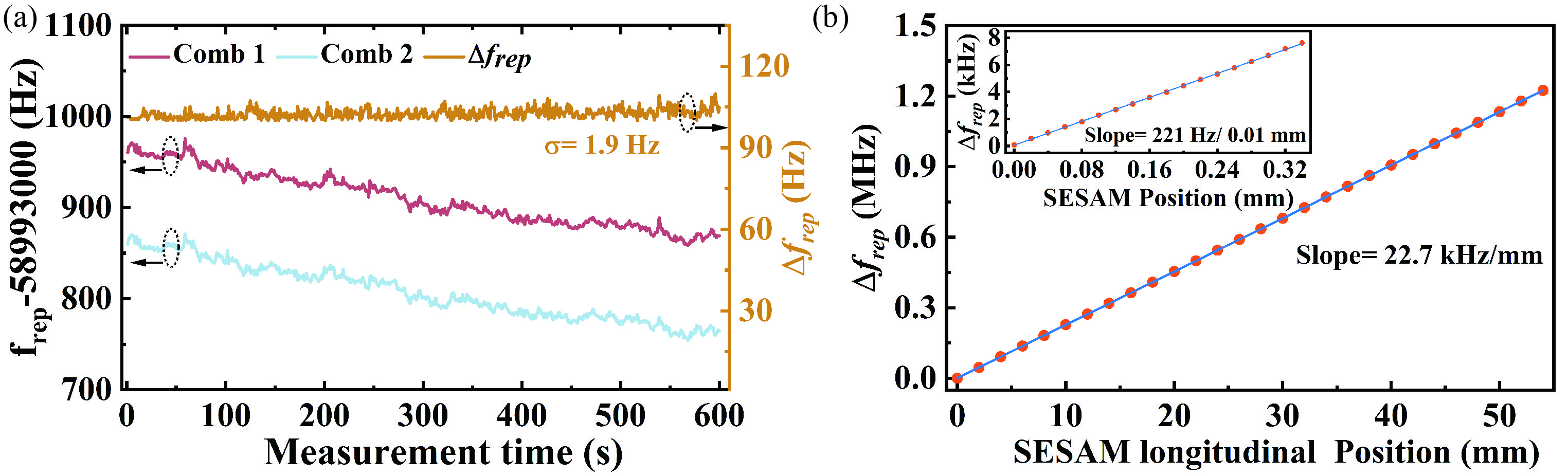}
	\caption{(a)Temporal evolution of the repetition rates and their difference. (b) Repetition rate difference versus the SESAM longitudinal position.}
\end{figure}

\section{Measurement of etalon transmission spectrum with free-running dual-comb laser}
\begin{figure}[ht!]
	\centering\includegraphics[width=13cm]{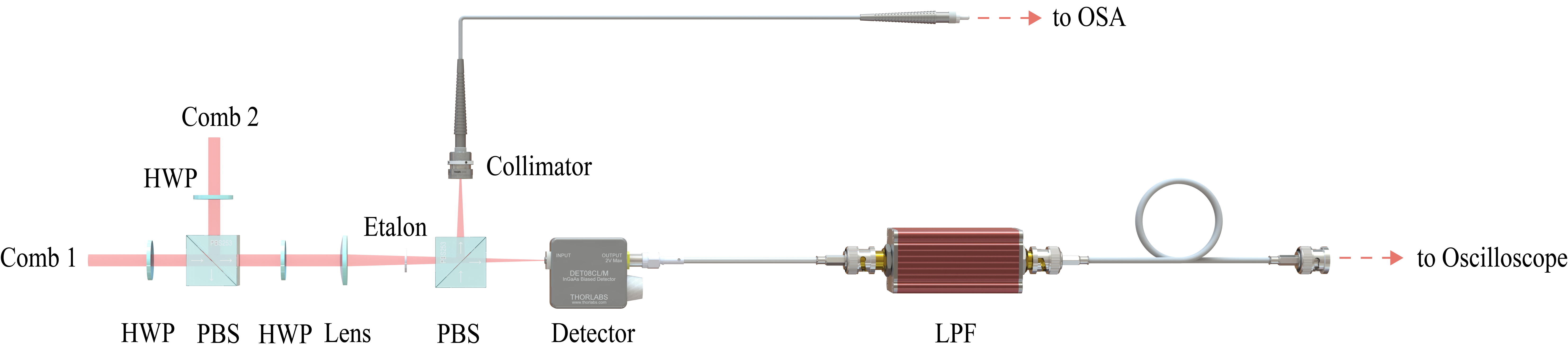}
	\caption{Interferogram measurment setup. HWP, halfwave plate; PBS, polarizing beamsplitter; LPF, low-pass filter; OSA, optical spectrum analyzer.}
\end{figure}

We measured the transmission spectrum of an etalon using a free-running dual-comb to investigate its potential for spectroscopic measurements. The RRD was set to 80 Hz. To capture the dual-comb spectrum, we constructed an interferogram measurement setup as illustrated in Fig. 4. The two laser beams from the combs were combined using a polarizing beam splitter (PBS) with orthogonal polarization and then passed through a half-wave plate and another PBS. As a result, the dual-comb laser was split into two collinear branches where both lasers had identical polarization directions and interfered with each other. One branch served as a reference and was directed towards an OSA (Yokogawa, AQ6370B). The other branch was detected by a fast photodetector (Thorlabs, DET08CL/M) followed by an 18 MHz low-pass filter and a digital phosphor oscilloscope (Tektronix, DPO7254C). During the measurement of the etalon's transmission spectrum, the etalon was positioned upstream of the second PBS. In this experiment, we utilized an uncoated YAG etalon that had a thickness of 250 $\upmu$m.
\begin{figure}[ht!]
	\centering\includegraphics[width=13cm]{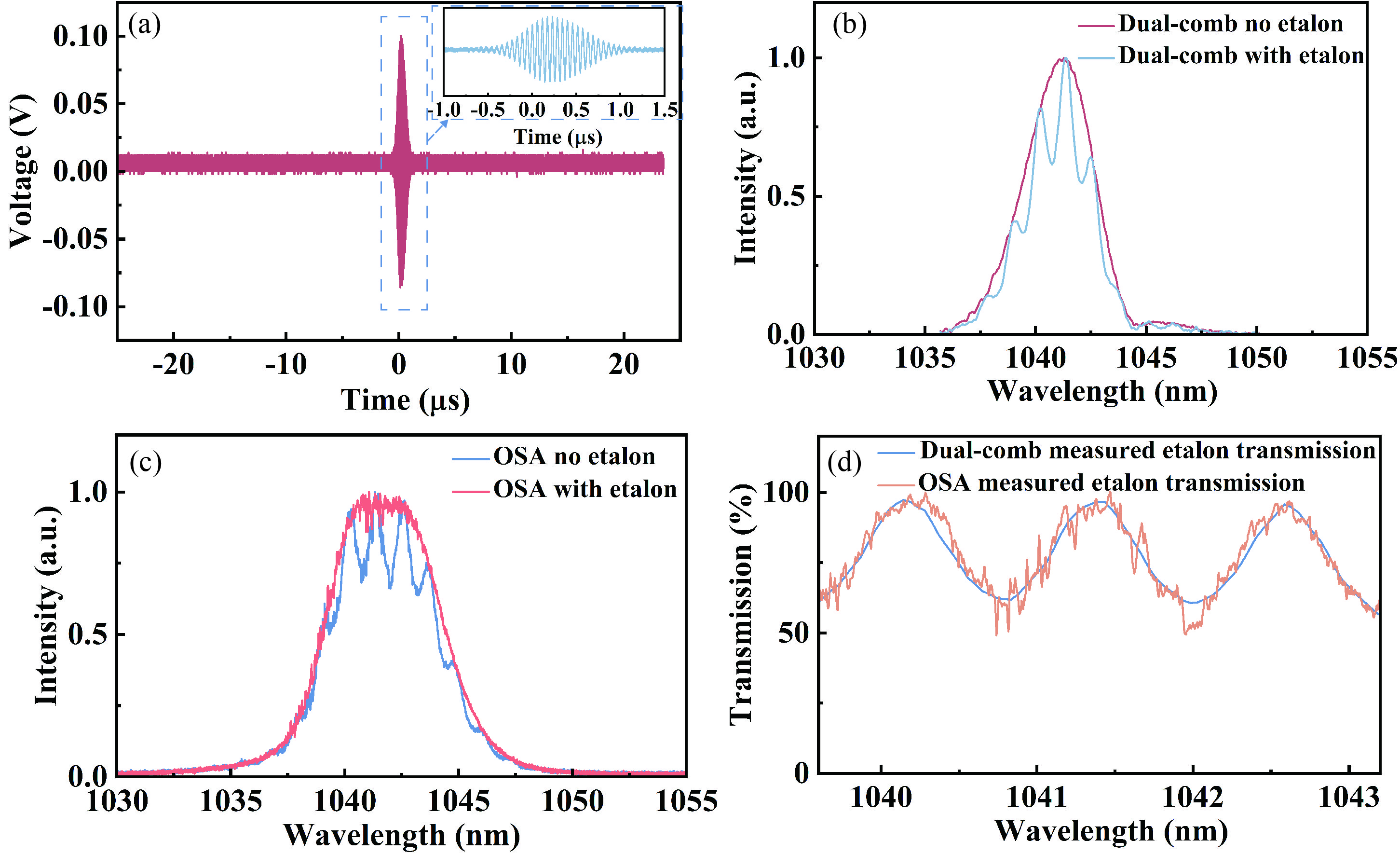}
	\caption{(a) Single-shot interferogram (inset: zoom-in of the interferogram in the blue box portion). (b) Comparison of the spectrum with and without etalon measured with dual-comb spectroscopy. (c) Comparison of the spectrum with and without etalon measured with the OSA. (d) Comparison of the etalon transmission spectrum measured with the dual-comb system and OSA.}
\end{figure}

To measure the dual-comb spectrum, we acquired a single-shot 50 $\upmu$s interferogram (Fig. 5(a)) and performed Fourier transformation into the frequency domain using the built-in Fast-Fourier-Transform (FFT) function of the oscilloscope (Fig. 5(b)). Simultaneously, the spectrum was recorded by an OSA (Fig. 5(c)). By comparing spectra obtained with or without an etalon, we derived the transmission spectrum (Fig. 5(d)). Although there was a slight difference between the spectra measured by our dual-comb system and those measured by OSA due to the partial overlap of comb spectra, excellent agreement in transmission spectra validates our dual-comb system's suitability for spectroscopic measurements.

\section{Measurement of the recovery time of SESAM based on ASOPS with free-running dual-comb  }
Given the direct output of asynchronous high-power femtosecond pulses with high mutual coherence, we employed this laser system in free-running to assess the recovery time of a SESAM using the ASOPS method. To measure the reflectivity of the commercially available BATOP SESAM, characterized by a modulation depth of 30$\%$, saturation fluence of 23 $\upmu$J/cm$^2$, and a recovery time constant of 0.7 ps, we established a non-collinear pump-probe setup as depicted in Fig. 6.
\begin{figure}[ht!]
	\centering\includegraphics[width=13cm]{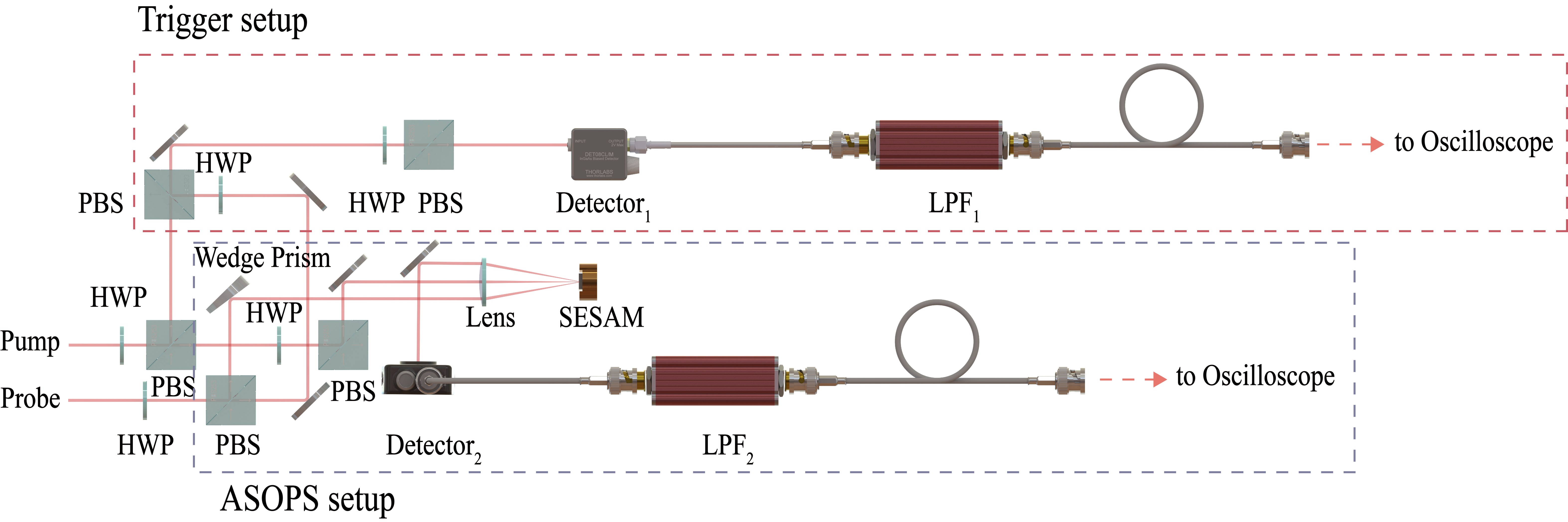}
	\caption{Setup of the ASOPS-based SESAM measurement. HWP, halfwave plates; PBS, polarizing beamsplitter; LPF, low-pass filter.}
\end{figure}
\begin{figure}[ht!]
	\centering\includegraphics[width=8cm]{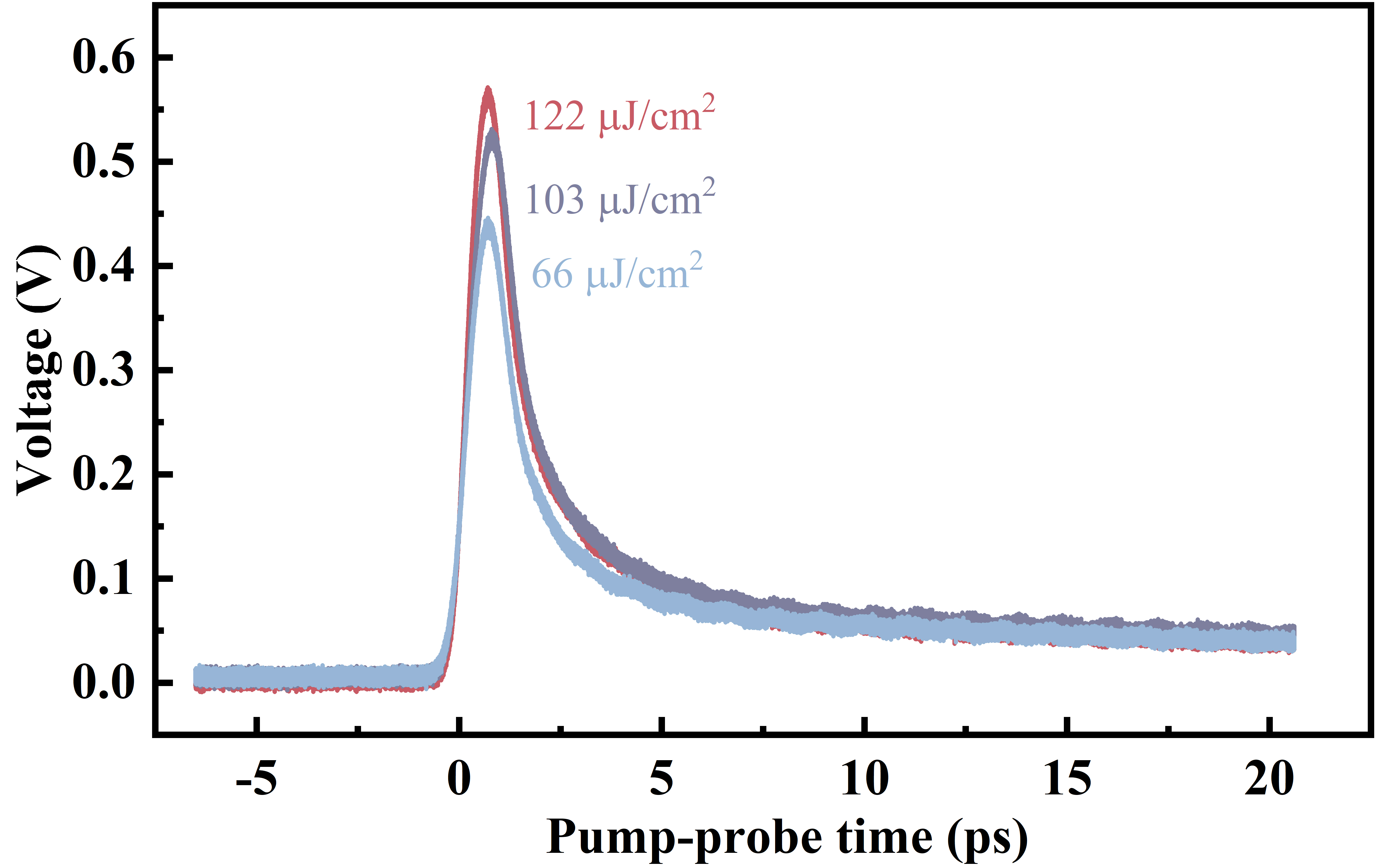}
	\caption{ASOPS-based SESAM measurement at different pump fluences.}
\end{figure}

In the ASOPS-based pump-probe setup, comb 2 served as the pump laser, while comb 1 was power-reduced to function as the probe laser. The SESAM surface was targeted by focusing the pump and probe beams using a 100-mm focal length lens with 1/e$^2$ diameters of 100 × 90 $\upmu$m$^2$ and 120 × 90 $\upmu$m$^2$, respectively (measured by Dataray Beam Profiler). The probe power was attenuated to 5 mW, whereas the pump power was sufficiently high to saturate the SESAM. A InGaAs photodetector (Thorlabs, DET20C2) detected the reflected probe laser signal which underwent low-pass filtering before being digitized and recorded on a digital phosphor oscilloscope (Tektronix, DPO7254C). Simultaneously, triggering of the oscilloscope occurred through an interferogram signal. For this experiment, we set RRD at 60 Hz resulting in an estimated step size for delay scanning of $\Delta$frep / frep$^2$ =17 fs; this value is smaller than the duration of the probe pulse. To convert to delay time scale, lab time scale is multiplied by a factor $\Delta$frep/ frep = (9.8 ×10$^5$)$^-$$^1$ . This ASOPS system enables rapid measurements that are unattainable with mechanically delayed scanning systems. Figure 7 illustrates measured results depicting reflectivity variation of SESAM under different pumping fluences. We extracted a fast recovery time constant of approximately 0.8 ps at a pump fluence level of 122 $\upmu$J/cm$^2$. These findings demonstrate that our free-running dual-comb system is well-suited for applications in pump-probe studies and its scanning speeds surpassing those achievable with any mechanical delay.

\section{Conclusion}
In summary, we have demonstrated a cavity-shared configuration with separated end mirrors for an all-solid-state mode-locked Yb:CALGO laser, resulting in the realization of a dual-comb oscillator with excellent controllability of the repetition rate difference and mutual coherence. The Yb:CALGO laser crystal is c-cut to ensure good isotropy for polarization. Two pairs of end mirrors and two TFPs angled at 45 degrees to the cavity axis are employed to separate the cross-polarized laser modes. Compared to the single cavity designs that share all elements, our design allows for more controllable repetition rate differences and easier cavity alignment. 

We achieved simultaneous self-starting modelocking at a center wavelength of 1043 nm (1040 nm) with average powers up to 1.07 W (1.14 W) and pulse duration around 400 fs for each comb. The repetition rates are approximately 59 MHz, and the RRD can be tuned from zero to 1.2 MHz without requiring realignment of the cavity or modification of other parameters. Common mode noises are effectively canceled out, resulting in minimal fluctuation of the RRD with a standard deviation of about 1.9 Hz over ten minutes, while repetition rates exhibit fluctuations as large as 90 Hz. 

Considering its output parameters, it exhibits promising potential for efficient nonlinear frequency conversion into spectral ranges that are pertinent to spectroscopy applications, such as the mid-infrared, terahertz, or ultraviolet regions. The power and pulse duration of the presented laser will be enhanced through improved mode matching between the laser modes and the pump laser, utilization of a pump laser with a more suitable wavelength, implementation of Kerr-lens mode-locking action, and incorporation of an optimized thin-film polarizer. Furthermore, although this dual-comb system is currently not actively locked for high-coherence operation, we anticipate higher performance in DCS or other precision-based measurements using dual-combs in future studies by applying feedback to the end mirrors.

\begin{backmatter}
\bmsection{Funding}
This work is supported by National Natural Science Foundation of China (62005215), Natural Science Foundation of Shaanxi Province(2024JC-YBMS-507), 
Natural Science Basic Research Program of Shaanxi Province (2019JCW‐03),
Science and Technology Program of Xi'an (202005YK01), Scientific Research Program of Shaanxi Education Department (23JK0690).

\bmsection{Disclosures}
The authors declare no conflicts of interest.
\end{backmatter}
\bibliography{Sample}






\end{document}